\RequirePackage{fix-cm}
\documentclass[natbib]{svjour3}     
\smartqed  % flush right qed marks, e.g. at end of proof
\usepackage{a4}
\usepackage{amsmath}
\usepackage{amssymb}
\usepackage{algorithm}
\usepackage{algorithmic}
\usepackage{wasysym}
\usepackage{graphicx}
\usepackage{url}
\usepackage{subfigure}
\usepackage{mathtools}
\usepackage{soul,color}
\sethlcolor{white}

\journalname{}

\begin{document}

\title{Competitive Fragmentation Modeling of ESI-MS/MS spectra for putative metabolite identification %\thanks{Grants or other notes
}

\titlerunning{CFM Modeling of ESI-MS/MS for putative Metabolite ID}        % if too long for running head

%\author{}
\author{Felicity Allen \and
        Russ Greiner \and
        % Yannick ??  \and
        David Wishart
}


\institute{
}
%F. Allen, R. Greiner and D. Wishart \at
             % Department of Computing Science, Athabasca Hall, University of Alberta, Edmonton Canada \\
%              %Fax: +123-45-678910\\
%              \email{felicity.allen@ualberta.ca}           %  \\
\vspace{-8em}
\date{}

\maketitle
\vspace{-1em}
\begin{abstract}
%\section{Motivation:}
Electrospray tandem mass spectrometry (ESI-MS/MS) is commonly used in high throughput metabolomics. 
One of the key obstacles to the effective use of this technology is the difficulty in interpreting measured spectra to accurately and efficiently identify metabolites. 
Traditional methods for automated metabolite identification compare the target MS or MS/MS spectrum to the spectra in a reference database, ranking candidates based on the closeness of the match. 
However the limited coverage of available databases has led to an interest in computational methods for predicting reference MS/MS spectra from chemical structures.

This work proposes a probabilistic generative model for the MS/MS fragmentation process, which we call Competitive Fragmentation Modeling (CFM), and a machine learning approach for learning parameters for this model from MS/MS data.
We show that CFM can be used in both a MS/MS spectrum prediction task (ie, predicting the mass spectrum from a chemical structure), and in a putative metabolite identification task (ranking possible structures for a target MS/MS spectrum). 

In the MS/MS spectrum prediction task, CFM shows significantly improved performance when compared to a full enumeration of all peaks corresponding to substructures of the molecule. 
In the metabolite identification task, CFM obtains substantially better rankings for the correct candidate than existing methods (MetFrag and FingerID) on tripeptide and metabolite data, when querying PubChem or KEGG for candidate structures of similar mass.
 
%\section{Availability:}
%Supplementary files containing test molecule lists and trained models are also available on that site.

%Insert your abstract here. Include keywords, PACS and mathematical
%subject classification numbers as needed.
\end{abstract}

\section{Introduction}
\label{intro}
%Metabolomics is a field of omics science that characterizes metabolites using high throughput technologies.
%Metabolites are all the low molecular weight ($<$1500 Da) chemicals found in cells, tissues and biofluids (\citep{Fiehn2002,Wishart2007}). They interact within complex regulatory networks to carry out many important life processes, such as growth, reproduction and signaling. 
%Understanding these complex biological processes may be key to the development of new biomonitoring, diagnostic or treatment technologies in areas such as agriculture and healthcare. 

%Liquid chromatography mass spectrometry (LC-MS) with electrospray ionization (ESI) is commonly used as an underlying platform for high-throughput metabolomics.

%In order to better understand such processes, researchers are seeking improved methods that measure metabolites in a high-throughput manner. 
%Many such methods use liquid chromatography mass spectrometry (LC-MS) with electrospray ionization (ESI) as their underlying platform. 
Liquid chromatography combined with Electrospray Ionisation Mass Spectrometry (ESI-MS) is one of the most frequently used approaches for conducting metabolomics experiments \citep{Dunn2005,Tautenhahn2012,Kind2010,Wishart2011}.
Collision-induced dissociation (CID) is usually employed within this procedure, intentionally fragmenting molecules into smaller parts to examine their structure. This is called MS/MS or tandem mass spectrometry. A significant bottleneck in such experiments is the interpretation of the resulting spectra to identify metabolites.

Widely used methods for putative metabolite identification \citep{Sumner2007}, using mass spectrometry, compare a collected MS or MS/MS spectrum for an unknown compound against a database containing reference MS or MS/MS spectra \citep{Stein1994, Scheubert2013, Tautenhahn2012}. 
Unfortunately, current reference databases are still fairly limited, especially in the case of ESI-MS/MS. At the time of writing, the public Human Metabolome Database \citep{Wishart2013} contains ESI-MS/MS data for around 800 compounds, which represents only a small fraction of the 40,468 known human metabolites it lists. The publicly available Metlin database \citep{Smith2005} provides ESI-MS/MS spectra for 11,209 of the 75,000 endogenous and exogenous metabolites it contains, although more than half of those spectra are for enumerated tripeptides. The public repository MassBank \citep{Horai2010} contains a more diverse dataset of 31,000 spectra collected on a variety of different instruments, including ESI-MS/MS spectra for approximately 2000 unique compounds. 
However, set against the more than 19 million chemical structures in the Pubchem Compound database \citep{Bolton2008}, an estimated 200,000 plant metabolites \citep{Fiehn2002}, or even the 32,801 manually annotated entries in the database of Chemical Entities of Biological Interest (ChEBI) \citep{Hastings2013}, we see that MS/MS coverage still falls far short of the vast number of known metabolites and molecules of interest. 

Consequently, there is substantial interest in finding alternative means for identifying metabolites for which no reference spectra are available \citep{Scheubert2013}.
For these cases, one approach to metabolite identification involves first predicting the MS or MS/MS spectrum for each candidate compound from its chemical structure \citep{Kokkonen2008, Wolf2010, Lindsay1980, Gasteiger1992}. 
The interpreter then uses these predicted spectra in place of reference spectra, and labels the target spectrum as the metabolite whose predicted spectrum is the closest match, according to some similarity criteria. 
A wide range of similarity criteria have been proposed, from weighted counts of the number of matching peaks \citep{Stein1994}, to more complex probability based measures \citep{Mylonas2009, Oberacher2009}.

The upshot of this predictive approach is that only a list of candidate molecules is needed, rather than a complete database of reference spectra. 
However, the restriction to a list of candidate molecules means that this approach still falls short of \emph{de novo} identification of 'unknown unknowns' \citep{Wishart2009}, -i.e. we cannot identify molecules not in the list. 
%However given a sufficiently comprehensive list of structures, or an exhaustive mechanism for generating potential structures, it may be possible to approach a comparable level of identification.

The concept of computer-based MS prediction has been around since the Dendral project in the 1960's, when investigators attempted to predict Electron Ionization (EI) mass spectra using early machine learning methods \citep{Lindsay1980}. 
More recent approaches to this problem have generally taken one of two forms: rule-based or combinatorial. 

Commercial packages, such as Mass Frontier (Thermo Scientific, \url{www.thermoscientific.com}), and MS Fragmenter (ACD Labs, \url{www.acdlabs.com}), are rule-based, using thousands of manually curated rules to predict fragmentations.  
Primarily developed for EI fragmentation, these packages have been extended for use with ESI.
This current work does not compare against these methods empirically, however in at least one study they have been found to have been out-performed by MetFrag \citep{Wolf2010}, to which we do compare.
\hl{MOLGEN-MS }\citep{Kerber2006}\hl{ also applies rule-based fragmentations in combination with an isotope-dependent matching criteria to rank candidate molecules for a given EI spectrum.} 
Another knowledge-based approach, called MASSIMO, combines chemical knowledge with data; using logistic regression to predict fragmentation probabilities for a particular class of EI fragmentations \citep{Gasteiger1992}. 
%While the authors of MASSIMO also propose a method for extracting more general fragmentation patterns from data, to the authors' knowledge, this method has never been successfully applied.

The other class of algorithms applies a combinatorial fragmentation procedure, enumerating all possible fragments of the original structure by systematically breaking bonds \citep{Hill2005, Kokkonen2008, Wolf2010}. 
First proposed by \citet{Hill2005}, this method has been incorporated into the freely available programs FiD \citep{Kokkonen2008} and MetFrag \citep{Wolf2010}. 
Both identify the given spectrum with the metabolite that has the most closely matching peaks via such a combinatorial fragmentation. 
These programs also employ several heuristics in their scoring protocols to emphasise the importance of more probable fragmentations. 
FiD uses an approximate measure of the dissociation energy of the broken bond, combined with a measure of the energy of the product ion. 
MetFrag incorporates a similar measure of bond energy combined with a bonus if the neutral loss formed is one of a common subset.

An alternative method, FingerID \citep{Heinonen2012}, takes advantage of the increasing number of available MS/MS spectra, by applying machine learning methods to this task. 
This program uses support vector machines (SVMs) to predict a chemical fingerprint directly from an MS/MS spectrum, and then searches for the metabolite that most closely matches that predicted fingerprint. 
\hl{For a more extensive review of existing computational methods in MS-based metabolite identification, see} \citet{Hufsky2014}.

%Like Heinonen \emph{et al.}, we think the metabolite identification problem may benefit from the application of machine learning methods. 
%However unlike their method, we choose not to discard the information obtained by the combinatorial methods. 
%Instead we use this enumeration of possible fragmentations as a starting point. 

The main problem with the current combinatorial methods is that, while they have very good recall, explaining most if not all peaks in each spectrum, they also have poor precision, predicting many more peaks than are actually observed. 
MetFrag and FiD attempt to address this problem by adding the heuristics described above. 
%However machine learning methods have been found to out-perform such hand-made rules in many applications \citep{Bishop2007}.
In our work, we investigate an alternative machine learning approach that aims to improve the precision of such combinatorial methods.

We propose a method for learning a generative model of the CID fragmentation process from data.
This model estimates the likelihood of any given fragmentation event occurring, thereby predicting those peaks that are most likely to be observed.
We hypothesise that increasing the precision of the predicted spectrum in this way will improve our system's ability to accurately identify metabolites.
\hl{In a similar spirit,} \citet{Kangas2012} \hl{proposed a machine learning approach for obtaining bond dissociation energies for lipids. Their method uses a different model and training paradigm which, to the authors' knowledge, has not yet been applied to general classes of metabolites.}

Section \ref{sec:cfm} provides details of our proposed model and the training method. Section \ref{sec:results} then reports the experimental results.
We will assume the reader knows the foundations of ESI MS/MS; for an introduction to this process, see \citet{HoffmanEdmondde;Stroobant2007}.

\section{\hl{Methods}}
\label{sec:cfm}

This section presents our model for the ESI-MS/MS CID fragmentation process, which we call Competitive Fragmentation Modeling (CFM), and a method for deriving parameters for this model from existing MS/MS data. 
Section~\ref{sec:SE} describes the simplest form of this method; Single Energy Competitive Fragmentation Modeling (SE-CFM). 
Section~\ref{CombinedEnergyModel} then presents an extension of this method, Combined Energy Competitive Fragmentation Modeling (CE-CFM), which aims to make better use of CID MS/MS spectra measured at different energy levels for the same compound. 

Windows executables and cross-platform source code and the trained models are freely available at
\url{http://sourceforge.net/projects/cfm-id/}.
A web server interface is also provided at \url{http://cfmid.wishartlab.com}. \hl{This provides access to the SE-CFM model trained on the Metlin Metabolite data as used in }Section~\ref{sec:results} \hl{, along with examples of predicted spectra.}

\subsection{Single Energy CFM (SE-CFM)}
\label{sec:SE}

\begin{figure*}
\begin{center}
\includegraphics[scale=0.5]{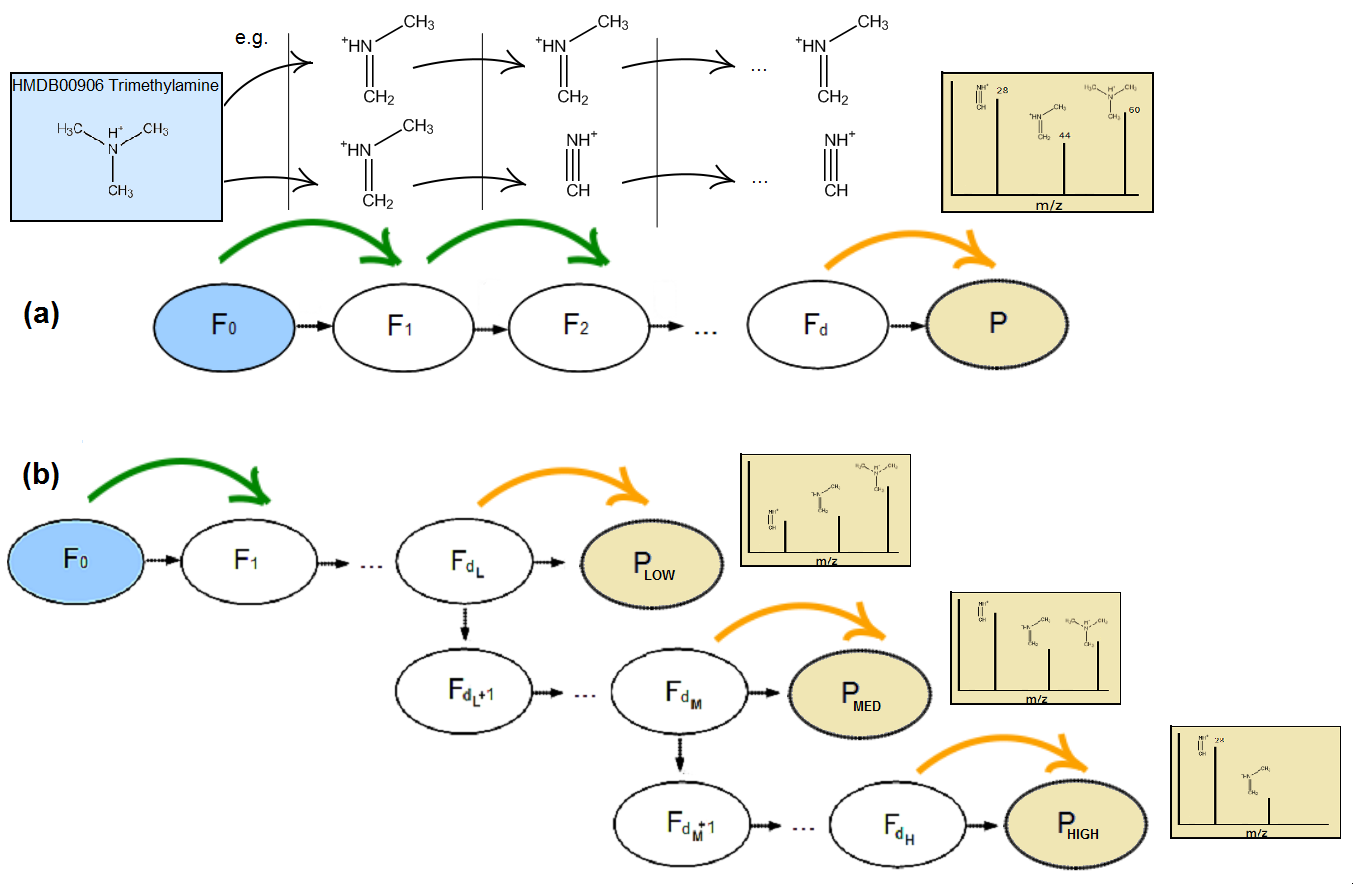}
\caption{(a) Single Energy Competitive Fragmentation Model (SE-CFM): a stochastic, Markov process of state transitions between charged fragments. (b) Combined Energy Competitive Fragmentation Model (CE-CFM): an extension of SE-CFM that combines information from multiple collision energy spectra into one model. }\label{model}
\label{fig:model}
\end{center}
\end{figure*}

In Single Energy CFM (SE-CFM), we model ESI-MS/MS fragmentation as a stochastic, homogeneous, Markov process \citep{Cappe2005} involving state transitions between charged fragments, as depicted in Figure~\ref{fig:model}(a).

More formally, the process is described by a fixed length sequence of discrete, random fragment states $F_{0}, F_{1}, \dots, F_{d}$, where each $F_{i}$ takes a value from the state space
$\mathcal{F} := \{f_{1},f_{2},\dots, f_{|\mathcal{F}|}\}$, 
the set of all possible fragments; this state space will be further described in Section~\ref{StateSpace}.
A transition model defines the probabilities that each fragment leads to another at one step in the process; see Section~\ref{TransitionModel}. 
An observation model maps the penultimate node $F_{d}$ to a peak $P$, which takes on a value in $\mathbb{R}$ that represents the m/z value of the peak to which the final fragment will contribute; see Section~\ref{ObservationModel}. 

SE-CFM is a latent variable model in which the only observed variables are the initial molecule $F_{0}$ and the output peak $P$; the fragments themselves are never directly observed. 
Each output $P$ adds only a small contribution to a single peak in the mass spectrum. In order to predict a complete mass spectrum, we can run the model forward multiple times to compute the marginal distribution of $P$.

\subsubsection{Fragment State Space}
\label{StateSpace}

We make the following assumptions about the CID fragmentation process. Further details for the motivations of each are provided below, but these generally involve a trade-off between accurately modeling the process and keeping the model computationally tractable.

\begin{enumerate}
\item All input molecules have a single positive charge and exist in their most common isotopic form. 
\item In a collision, each molecule will break into two fragments.
\item No mass or charge is lost. One of the two fragments must have a single positive charge and the other must be neutral. Combined, the two must contain all the components of the original charged molecule, i.e. all the atoms and electrons.
\item No further sigma bonds can be removed or added during a break, except those connecting hydrogens --i.e. the edges in the molecular graph must remain the same.
\item Rearrangement of pi bonds is allowed and hydrogen atoms may move anywhere in the two resulting fragments, on the condition that both fragments satisfy all valence rules, and standard bond limitations are met --e.g. no bond orders higher than triple.
\item The even electron rule is always satisfied --i.e. no radicals.
\end{enumerate}

Assumption 1 is reasonable as we assume that the first phase of MS/MS successfully restricts the mass range of interest to include only the [M+H]$^{+}$ precursor ion containing the most abundant isotopes. Since this ion has only a single positive charge, we can safely assume that no multiply-charged ions will be formed in the subsequent MS2 phase. Ensuring that valid [M+H]$^{+}$ precursor ions are selected in MS1 is beyond the scope of this work; see \citet{Katajamaa2007} for a summary of MS1 data processing methods.

Assumptions 2, 4 and 6 do not necessarily hold in real-world spectra \citep{Galezowska2013, Levsen2007}. However including them substantially reduces the branching factor of the fragment enumeration, making the computations feasible. Since these assumptions do appear to hold in the vast majority of cases, we expect that including them should have minimal negative impact on the experimental results. 
Note that most 3-way fragmentations can be modeled by two sequential, 2-way fragmentations, so including Assumption 2 should not impact our ability to model most fragmentation events.
Assumption 5 allows for McLafferty Rearrangement and other known fragmentation mechanisms \citep{McLafferty1993}. 

Our method for enumerating fragments is similar in principle to the combinatorial approach used in MetFrag and FiD \citep{Wolf2010, Kokkonen2008}, with some additional checks to enforce the above assumptions.  
We systematically break all non-ring bonds in the molecule (excluding those connecting to hydrogens) and all pairs of bonds within each ring. We do this one break at a time, enumerating a subset of fragments with all possible masses that may form after each break, \hl{allowing for hydrogen rearrangements}. This subset is found by determining the number of additional electrons that can be allocated to either side of the break using integer linear programming \hl{to enforce bond constraints} --e.g. breaking the middle bond in CCC[CH4+] (SMILES format) gives \hl{possible} fragments C=[CH3+] (mass=29.04Da, loss CC) and C[CH4+] (mass=31.05Da, loss C=C), \hl{whereas it is not possible to break the triple bond in C\#[CH2+] because there is nowhere for the electrons from the bond to go.}

\begin{figure}
\begin{center}
\includegraphics[scale=0.45]{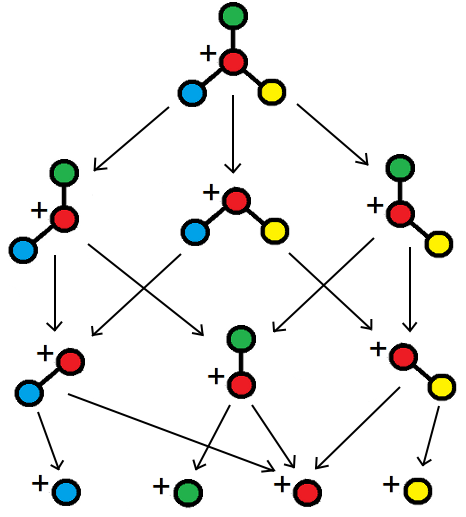}
\end{center}
\caption{An abstract example of a fragmentation graph, showing a directed acyclic graph of all possible ways in which a particular charged molecule may break to produce smaller charged fragments.}
\label{fig:fragmentationgraph}
\vspace{-1em}
\end{figure}

The fragmentation procedure is applied recursively on all the produced fragments, to a maximum depth. 
The result is a directed acyclic graph (DAG) containing all possible charged fragments that may be generated from that molecule. 
An abstract example of such a fragmentation graph is provided in Figure~\ref{fig:fragmentationgraph}. 
Note that for each break, one of the two produced fragments will have no charge. Since it is not possible for a mass spectrometer to detect neutral molecules, 
we do not explicitly include the neutral fragments in the resulting graph, nor do we recur on their possible breaks. 
However neutral loss information may be included on the edges of the graph, indicating how a particular charged fragment was determined.
\hl{This representation of the fragmentation possibilities as a DAG is similar to that proposed by} \citet{Bocker2008} \hl{with the exception that their nodes contain molecular formulae rather than structures for the ions.}

\subsubsection{Transition Model}
\label{TransitionModel}

Our parametrized transition model assigns a conditional probability to each fragment given the previous fragment in the sequence $F_{0}$,$F_{1}$,\dots,$F_{d}$. 
\hl{Recall that $F_{t}$ denotes the random fragment state at time $t$, whereas $f_{i}$ denotes the $i$th fragment in the space of all fragments.} 
In the case where $f_{i}$ has $f_{j}$ as a possible child fragment in a fragmentation graph, our model assigns a positive probability to the transition from $F_{t} = f_{i}$ to $F_{t+1} = f_{j}$. 
Furthermore, self-transitions are always allowed, i.e. the probability of transitioning from $F_{t} = f_{i}$ to $F_{t+1} = f_{i}$ is always positive (for the same $f_{i}$). 
We assign 0 probability to all other transitions, i.e. those that are not self-transitions, and that do not exist within any fragmentation graph. 

Although the set of possible charged fragments $\mathcal{F}$ is large, the subset of child fragments originating from any particular fragment is relatively small. 
For example, the requirement that a feasible child fragment must contain a subset of the atoms in the parent fragment rules out many possibilities.  
Consequently most transitions will be assigned a probability of 0. 
Note that the assigned probabilities of all transitions originating at a particular fragment, including the self-transition, must sum to one.

We now discuss how we parametrize our transition model.
A natural parametrization would be to use a transition matrix containing a separate parameter for every possible fragmentation $f_{i}\rightarrow f_{j}$.
Unfortunately, we lack sufficient data to learn parameters for every individual fragmentation in this manner. 
Instead, we look for methods that can generalize by exploiting the tendency of similar molecules to break in similar ways.

\subsubsection{Break Tendency}
 
\begin{figure}[!tpb]
\begin{center}
\includegraphics[scale=0.7]{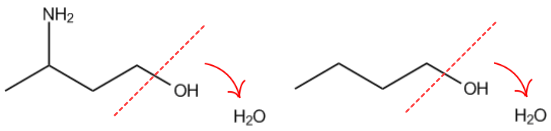}
\caption{\label{fig:twobreaks}Two similar breaks, both resulting in an H$_{2}$O neutral loss. 
The right case should be assigned a higher probability, as in the left case, the NH$_{3}$ is also likely to break away, reducing the probability of the H$_{2}$O loss.}
\end{center}
\end{figure}

We introduce the notion of \emph{break tendency}, which we represent by a value $\theta \in \mathbb{R}$ for each possible fragmentation $f_{i}\rightarrow f_{j}$ that models how likely a particular break is to occur. 
Those fragmentations that are more likely to occur are assigned a higher break tendency value, and those that are less likely are given lower values.  
We then employ a softmax function to map the break tendencies for all breaks involving a particular parent fragment to probabilities, as defined in Equation \ref{rhoEquation} below. 
This has the effect of capturing the competition that occurs between different possible breaks within the same molecule.
For example, consider the two fragmentations in Figure \ref{fig:twobreaks}.
Here, although both fragmentations involve an H$_{2}$O neutral loss, in the left-hand case, the H$_{2}$O loss must compete with the loss of an ammonia group, whereas in the right hand case, it does not.
Hence our model might assign an equal break tendency to both cases, but this would still result in a lower probability of fragmentation in the former case, due to the competing ammonia.
 
We model the probability of a particular break $f_{i} \rightarrow f_{j}$ occurring as a function of its break tendency value $\theta_{i,j}$ and that of all other competing breaks from the same parent, as follows:
\begin{equation}
\label{rhoEquation}
\rho(f_{i},f_{j}) = 
\begin{dcases}
\frac{\exp{\theta_{i,j}}}{1 + \sum\limits_{k}\exp{\theta_{i,k}}} & : f_{i}\neq f_{j} \text{ and } f_{i} \rightarrow f_{j} \text{ is possible} \\
\frac{1}{1 + \sum\limits_{k}\exp{\theta_{i,k}}} & : f_{i}=f_{j}\\
0 & :  f_{i} \rightarrow f_{j} \text{ is not possible}
\end{dcases}
\end{equation}
where the sums iterate over all $k$ for which $f_{i} \rightarrow f_{k}$ is possible. 

Since the break tendency is a relative measure, it makes sense to tie it to some reference point. 
For the purposes of this model, we have assigned the break tendency for a self-transition (i.e. no break occuring) to $\theta_{i,i} = 0$, which gives $\exp{\theta_{i,i}}=1$ as shown in (\ref{rhoEquation}). 

\paragraph{Incorporating Chemical Features} 
We need to compute $\theta_{i,j}$ for $i\neq\nobreak j$. 
To do this we first define a binary feature vector $\Phi_{i,j}$ to describe the characteristics of a given break $f_{i}\rightarrow f_{j}$. 
Such features might include the presence of a particular atom adjacent to the broken bond, or the formation of a specific neutral loss molecule --e.g. see Section~\ref{sec:ChemicalFeatures}.
We then use these features to assign a break tendency value using a linear function parameterized by a vector of weights $w \in \mathbb{R}^{n}$ --i.e. $\theta_{i,j} := w^{T}\Phi_{i,j}$. This can then be substituted into (\ref{rhoEquation}) to generate the probability of transition $f_{i} \rightarrow f_{j}$.  The first feature of $\Phi_{i,j}$ is a bias term, set to 1 for all breaks.
Note that the vector $w$ constitutes the parameters of the CFM model that we will be learning.

\subsubsection{Observation Model}
\label{ObservationModel}

\newcommand{\mass}{\text{mass}}

We model the conditional probability of $P$ using a narrow Gaussian distribution centred around the mass\footnote{Although mass spectrometry measures mass over charge, we assume charge is always 1 (see Assumption 1 in Section~\ref{StateSpace}) and hence can just use the mass here.} of $F_{d}$, i.e. $P|F_{d} \sim \mathcal{N}( \mass(F_{d}), \sigma^{2} )$. 
The value for $\sigma$ can be set according to the mass accuracy of the mass spectrometer used. 
So, we define this observation function to be the following
\begin{equation}
\label{obsEquation}
g(m,F_{d};\sigma) = \frac{1}{\sigma\sqrt{2\pi}}\exp \left\{-\frac{1}{2} \left(\frac{m - \mass(F_{d})}{\sigma}\right )^{2} \right\}.
\end{equation}
Our investigation (see supplementary data) of the mass error of the precursor ions in the Metlin metabolite data used in Section~\ref{sec:results} found that the distribution of mass errors 
had a mean offset of approximately 1~ppm, and a narrower shape than a Gaussian distribution.
However, in order to model a more general mass error, not specific to a particular instrument or set of empirical data, we think the Gaussian distribution is a reasonable approach.

\subsubsection{Selecting Parameter Values}
\label{SimpleSelectingParams}

Our system estimates the values for the parameters $w$ of the proposed model by applying a training procedure to a set of molecules $\mathcal{X} = \{x_{1},x_{2},\dots,x_{|\mathcal{X}|}\}$, for which we have both the chemical structure and a measured MS/MS spectrum. 

For the purposes of this work, we assume we have a measured low, medium and high energy CID MS/MS spectrum for each molecule, which we denote  
$S(x) = $( $s_{L}(x)$, $s_{M}(x)$, $s_{H}(x)$)$ \forall x\in\mathcal{X}.$ 
Each spectrum is further defined to be a set of peaks, where each peak is a pair $(m,h)$, composed of a mass $m \in \mathbb{R}$ and a height (or intensity) $h \in [0,100] \subset \mathbb{R}$. Note that each spectrum is normalized, such that the peak heights sum to 100.

For this single energy version of the model, we derive parameters for a completely separate model for each of the three energy levels, using data from that level only.
Note that if we had data for only one energy level, we could use this method to train a model using just that energy. 
However Section~\ref{CombinedEnergyModel} will extend this model to combine the three energy spectra for use in a single model. 
Until then, we will use $s(x)$ to denote whichever of $s_{L}(x)$, $s_{M}(x)$ or $s_{H}(x)$ we are currently considering.

\paragraph{Maximum Likelihood}
We use a Maximum Likelihood approach for parameter estimation.
The likelihood of the data $\mathcal{X}$, given the parameters $w$,  
and incorporating the previously defined transition function $\rho$ and observation function $g$, is given by
\begin{equation}
%\begin{align*}
\mathcal{L}(w , \mathcal{X}) 
 \hspace{4pt}=\prod\limits_{x \in \mathcal{X}}\prod\limits_{(m,h) \in s(x)}\hspace{-4pt}\Big(\hspace{-3pt}\sum\limits_{F_{1}\in C'(x)}\hspace{-10pt} \rho( x,F_{1}; w)\hspace{-10pt} \sum\limits_{F_{2}\in C'(F_{1})}\hspace{-12pt} \rho( F_{1},F_{2}; w)   
 \dots\hspace{-16pt}\sum\limits_{F_{d}\in C'(F_{d\hspace{-1pt}-\hspace{-1pt}1})}\hspace{-14pt}\rho( F_{d\hspace{-1pt}-\hspace{-1pt}1},F_{d}; w)\text{ }g(m,F_{d}; \sigma)\Big)^{h}
%\end{align*}
\end{equation}
where $C(f_{i})$ denotes the children of $f_{i}$ in all fragmentation graphs containing it, and $C'(f_{i}) = \{f_{i}\} \cup C(f_{i})$.

However we are unable to maximize this function in closed form. 
Instead we use the iterative Expectation Maximization \citep{Dempster1977} technique.

\paragraph{Expectation Maximization (EM)}

In the E-step, the expected log likelihood expression is given by
\begin{align}
Q(w^{t},w^{t\hspace{-1pt}-\hspace{-1pt}1}\,|\,\mathcal{X})
&\quad=\quad\mathbb{E}_{w^{t\hspace{-1pt}-\hspace{-1pt}1}}\big(\log{\mathcal{L}(w^{t},\mathcal{X})}\big) \\
&\quad=\quad\sum\limits_{F_{1}}\hspace{-6pt}\dots\hspace{-4pt}\sum\limits_{F_{d}}\Pr\big(F_{1} \dots F_{d} \,|\, \mathcal{X}; w^{t\hspace{-1pt}-\hspace{-1pt}1}\big)\log{\mathcal{L}(w^{t},\mathcal{X})},
\end{align}
where $w^{t}$ denotes the values for $w$ on the $t$-th iteration. Substituting (\ref{rhoEquation}) and (\ref{obsEquation}) into the above and re-arranging in terms of all possible fragment pairs gives
\begin{align}
Q(w^{t},w^{t\hspace{-1pt}-\hspace{-1pt}1}\,|\,\mathcal{X})\quad = \hspace{-5pt}\sum\limits_{(f_{i},f_{j}) \in \mathcal{F}\times\mathcal{F}} \hspace{-12pt} \nu_{w^{t\hspace{-1pt}-\hspace{-1pt}1}}(f_{i},f_{j},\mathcal{X})\log\rho(f_{i},f_{j};w^{t}) + K
\end{align}
where 
\begin{equation*}
\label{nuEquation}
\nu_{w^{t\hspace{-1pt}-\hspace{-1pt}1}}(f_{i},f_{j},\mathcal{X}) \quad=\quad \sum\limits_{d'=1}^{d}\eta_{w^{t\hspace{-1pt}-\hspace{-1pt}1}}^{d'}(f_{i},f_{j},\mathcal{X}),
\end{equation*}
\begin{equation*}
\eta_{w^{t\hspace{-1pt}-\hspace{-1pt}1}}^{d}(f_{i},f_{j},\mathcal{X})\quad= \hspace{-15pt}\sum\limits_{\{(m,h) \in s(x):x \in \mathcal{X}\}}\hspace{-24pt}h \Pr\big(F_{d\hspace{-1pt}-\hspace{-1pt}1}\hspace{-3pt}=\hspace{-3pt}f_{i},F_{d}\hspace{-3pt}=\hspace{-3pt}f_{j} \,|\, F_{0}\hspace{-3pt}=\hspace{-3pt}x, P\hspace{-3pt}=\hspace{-3pt}m; w^{t\hspace{-1pt}-\hspace{-1pt}1}\big)
\end{equation*}
and
\begin{equation*}
K \quad=\quad \sum\limits_{F_{d}}\Pr(F_{d}\,|\, \mathcal{X}; w^{t\hspace{-1pt}-\hspace{-1pt}1})\log\Pr(P=m \,|\, F_{d}).
\end{equation*}

In the M-Step, we look for the $w^{t}$ that maximizes the above expression of $Q$. Noting that $K$ is independent of $w^{t}$ and denoting the $l$th component of $w$ as $w_{l}$, 
\begin{equation}
\label{gradEquation}
\frac{\partial Q}{\partial w_{l}} \quad= \sum\limits_{(f_{i},f_{j}) \in \mathcal{F}\times\mathcal{F}} \hspace{-15pt}\nu_{w^{t\hspace{-1pt}-\hspace{-1pt}1}}(f_{i},f_{j},\mathcal{X})\Big(\mathbb{I}[f_{i}\hspace{-3pt}\neq\hspace{-3pt}f_{j}]\Phi_{i,j}^{l} - \hspace{-8pt}\sum\limits_{k\in C(f_{i})}\hspace{-8pt}\Phi_{i,k}^{l}\rho(f_{i},f_{k};w)\Big)
\end{equation}
where $\Phi_{i,k}^{l}$ denotes the $l$th component of the feature vector $\Phi_{i,k}$ and $\mathbb{I}[.]$ is the indicator function.

This does not permit a simple closed-form solution for $w$. However $Q(w^{t},w^{t\hspace{-1pt}-\hspace{-1pt}1}\,|\,\mathcal{X})$  is concave in $w^{t}$, so settings for $w^{t}$ can be found using gradient ascent. 
Values for the joint probabilities in the $\eta_{w^{t\hspace{-1pt}-\hspace{-1pt}1}}^{d}$ terms can be computed efficiently using the junction tree algorithm \citep{Koller2009}. 

We also add an $\ell_{2}$ regularizer on the values of $w$ to $Q$ (excluding the bias term). 
This has the effect of discouraging overfitting by encouraging the parameters to remain close to zero.

\subsection{Combined Energy CFM}
\label{CombinedEnergyModel}

MS/MS spectra are often collected at multiple collision energies for the same molecule. 
Increasing the collision energy usually causes more fragmentation events to occur.
This means that fragments appearing in the medium and high energy spectra are almost always descendants 
of those that appear in the low and medium energy spectra, respectively.
So the existence of a peak in the medium energy spectrum may help to differentiate between explanations for a related peak in the low or high energy spectra. 

For this reason, we also assessed an additional model, Combined Energy CFM (CE-CFM), which extends the SE-CFM concept by combining information from multiple energies as shown in Fig.~\ref{model}~(b). P{\tiny LOW}, P{\tiny MED} and P{\tiny HIGH} each represent a peak from the low, medium and high energy spectrum respectively.
%\begin{figure}[!tpb]
%\begin{center}
%\includegraphics[scale=0.42]{combined_model_rev.png}
%\caption{Combined Energy Competitive Fragmentation Model (CE-CFM) combines information from multiple collision energy spectra into one model. P{\tiny LOW}, P{\tiny MED} and P{\tiny HIGH} each represent a peak from the low, medium and high energy spectrum respectively. }
%\label{combined}
%\end{center}
%\vspace{-1.25em}
%\end{figure} 
The fragment states, transition rules and the observation model are all the same here as for SE-CFM. 
The main difference now is that the homogeneity assumption is relaxed so that separate transition likelihoods can be learned for each energy block --i.e., $F_{0}$ to $F_{d_{L}}$, $F_{d_{L}}$ to $F_{d_{M}}$ and $F_{d_{M}}$ to $F_{d_{H}}$, where $d_{L}$, $d_{M}$ and  $d_{H}$ denote the fragmentation depths of the low, medium and high energy spectra respectively. This results in separate parameter values for each energy, denoted respectively as $w_{L}$, $w_{M}$ and $w_{H}$. 
The complete parameter set for this model thus becomes $w = w_{L}\cup w_{M}\cup w_{H}$. 

We can again use a Maximum Likelihood approach to parameter estimation based on the EM algorithm. This approach deviates from the SE-CFM method only as follows:
\begin{itemize}
\item For each energy level, (\ref{gradEquation}) is computed separately, restricting the $\nu_{w^{t\hspace{-1pt}-\hspace{-1pt}1}}$ terms to relevant parts of the model --e.g. $d'$ would sum from $d_{L}\hspace{-2pt}+\hspace{-2pt}1$ to $d_{M}$ when computing the gradients for $w_{M}$, and from $d_{M+1}$ to $d_{H}$ when computing gradients for $w_{H}$.
\item The computation of the $\eta^{d}_{w_{t\hspace{-1pt}-\hspace{-1pt}1}}$ terms combines evidence from the full set of three spectra $S(x)$. In SE-CFM, we apply one spectrum at a time, effectively sampling from a distribution over the peaks from each observed spectra. In this extended model we cannot do this because we do not have a full joint distribution over the peaks, but rather we only have marginal distributions corresponding to each spectrum. 
The standard inference algorithms --e.g. the junction tree algorithm, do not allow us to deal with observations that are marginal distributions rather than single values. Instead we use the Iterative Proportional Fitting Procedure (IPFP) \citep{Stephan1940}, with minor modifications to better handle cases where the spectra are inconsistent (not simultaneously achievable under any joint distribution). These modifications reassign the target spectra to be the average of those encountered when the algorithm oscillates in such circumstances.
\end{itemize}

\section{Experimental Results}
\label{sec:results}

In this section we present results using the above described SE-CFM ($d$=2) and CE-CFM ($d_{L}$=2, $d_{M}$=4, $d_{H}$=6) methods, on a spectrum prediction task, and then in a metabolite identification task. 

\subsection{Data}
We used the Metlin database \citep{Smith2005},
separated into two sets (see description below) each containing positive mode, ESI-MS/MS spectra from a 6510 Q-TOF (Agilent Technologies) mass spectrometer, measured at three different collision energies: 10V, 20V and 40V, which we assign to be low, medium and high energy respectively.
Each set was randomly divided into 10 groups for use within a 10-fold cross validation framework.
\begin{enumerate}

\item \textbf{Tripeptides}: The Metlin database contains data for over 4000 enumerated tripeptides. We randomly selected 2000 of these molecules, then omitted 15 that had four or more rings due to computational resource concerns, leaving 1985 remaining in the set.
Fragmentation patterns in peptides are reasonably well understood \citep{Papayannopoulos1995, Paizs2005}, leading to effective algorithms for identifying peptides from their ESI MS/MS data --e.g. \citep{Pappin1999,Eng1994,Ma2003}.  
However, we think that the size of this dataset, and the fact that it contains so many similar yet different molecules, make it an interesting test case for our algorithms. 

\item \textbf{Metlin Metabolites}: We use a set of 1491 non-peptide metabolites from the Metlin database.  These are a more diverse set covering a much wider range of molecules. An initial set of 1500 were selected randomly. Nine were then excluded because they were so much larger than the other molecules (over 1000 Da), such that their fragmentation graphs could not be computed in a reasonable amount of time.
\end{enumerate}

We also used an additional small validation set, selected because they were measured on a similar mass spectrometer, an Agilent 6520 Q-TOF, but in a different laboratory. These were taken from the MassBank database \citep{Horai2010}. All testing with this set used a model trained for the first cross-fold set of the Metlin metabolite data ($\sim 90\%$ of the data).
\begin{enumerate}
\setcounter{enumi}{2}
\item \textbf{MassBank Metabolites}: This set contains 192 metabolites taken from the Washington State University submission to the MassBank database. All molecules from this submission were included that had MS2 spectra with collision energies 10V, 20V and 40V, in order to provide a good match with the Metlin data.
\end{enumerate}

Files containing test molecule lists and assigned cross validation groups are provided as supplementary data. %\url{sourceforge.net/projects/cfm-id/}.

\subsection{Chemical Features}
\label{sec:ChemicalFeatures}

\begin{figure}[!tpb]
\begin{center}
\includegraphics[scale=0.7]{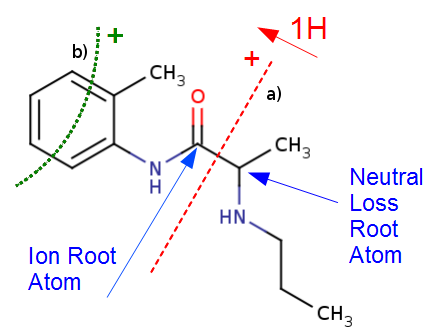}
\caption{Two example fragmentations. a) A non-ring break for which the ion and neutral loss root atoms are labeled. The 1H indicates the movement of a hydrogen to the ion side  (marked with a +) from the neutral loss side. b) A ring break for a single aromatic ring of size 6, in which the distance between the broken bonds is 3.  }
\label{fig:breakLabel}
\end{center}
\vspace{-2em}
\end{figure}

The chemical features used in these experiments were as follows. 
Note that the terms \emph{ion root atom} and \emph{neutral loss (NL) root atom} refer to the atoms connected to the broken bond(s) on the ion and neutral loss sides respectively --cf., Fig.~\ref{fig:breakLabel}.
\begin{itemize}
\item \emph{Break Atom Pair}: Indicators for the pair of ion and neutral loss root atoms, each from \{C,N,O,P,S, other\}, included separately for those in a non-ring break vs those in a ring break --e.g. Fig.~\ref{fig:breakLabel}a): would be non-ring C-C. (72 features)
\item \emph{Ion and NL Root Paths} Indicators for all paths of length 2 and 3 starting at the respective root atoms and stepping away from the break. Each is an ordered double or triple from \{C,N,O,P,S,other\}, taken separately for rings and non-rings. Two more features indicate no paths of length 2 and 3 respectively --e.g. Fig.~\ref{fig:breakLabel}a): the ion root paths are C-O, C-N and C-N-C. (2020 features).
\item \emph{Gasteiger Charges}: Indicators for the quantised pair of Gasteiger charges \citep{Gasteiger1980} for the ion and NL root atoms in the original unbroken molecule. (288 features)
\item \emph{Hydrogen Movement}: Indicator for how many hydrogens switched sides of the break and in which direction --i.e. ion to NL (-) or NL to ion(+) \{0,$\pm1$,$\pm2$,$\pm3$,$\pm4$,other\}. (10 features)
\item \emph{Ring Features}: Properties of a broken ring. Aromatic or not? Multiple ring system? Size \{3,4,5,6, other\}? Distance between the broken bonds \{1,2,3,4+\}? --e.g. Fig.~\ref{fig:breakLabel}b) is a break of a single aromatic ring of size 6 at distance 3. (12 features).
\end{itemize}

Of these 2402 features, few take non-zero values for any given break. 
Many are never encountered in our data set, in which case their corresponding parameters are set immediately to 0.
We also append \emph{Quadratic Features}, containing all 2,881,200 pair-wise combinations of the above features, excluding the additional bias term. Again, most are never encountered, so their parameters are set to 0.

\subsection{Spectrum Prediction}
\label{sec:SpectrumPrediction}

For each cross validation fold, and the MassBank validation set, a model (trained as above), was used to predict a low, medium and high energy spectra for each molecule in the test set. 
The \hl{model is run forward and the} resulting marginal distributions for the peak variables are a mixture of Gaussian distributions. 
We take the means and weights of these Gaussians as our peak mass and intensity values. 
Since all fragments in the fragmentation graph of a molecule have non-zero probabilities in the marginal distribution, it is necessary to place a cut-off on the intensity values to select only the most likely peaks. 
Here, we use a post-processing step that removes peaks with low probability, keeping as many of the highest peaks as required to form at least 80\% of the total intensity sum.  
We also set limits on the number of selected peaks to be at least 5 and at most 30. This ensures that more peaks are included than just the precursor ion, and also prevents spectra occurring that have large numbers of very small peaks. These values were selected arbitrarily, but post-analysis suggests that they are reasonable (see supplementary data). 
When matching peaks we use a mass tolerance set to the larger of 10~ppm and 0.01~Da (depending on the peak mass), and set the observation parameter $\sigma$ to be one third of this value.
\hl{No additional processing was done for the experimental spectra.}

\paragraph{Metrics}
We consider a peak in the predicted MS/MS spectrum $s_{P}$ to match a peak in the measured MS/MS spectrum $s_{M}$ if their masses are within the mass tolerance above.
We use the following metrics:

\begin{enumerate}
\item \textbf{Weighted Recall}: The percentage of the total peak intensity in the measured spectrum with a matching peak in the predicted spectrum:
	$ 100 \times \hspace{-10pt}\sum\limits_{(m,h)\in s_{M}} \hspace{-10pt}h \cdot \mathbb{I}[(m,h) \in s_{P}] \hspace{2pt}\div \hspace{-10pt}\sum\limits_{(m,h)\in s_{M}}\hspace{-12pt}h$.
\item \textbf{Weighted Precision}: The percentage of the total peak intensity in the predicted spectrum with a matching peak in the measured spectrum:
	$ 100 \times \hspace{-10pt}\sum\limits_{(m,h)\in s_{P}} \hspace{-10pt}h \cdot \mathbb{I}[(m,h) \in s_{M}] \hspace{2pt}\div \hspace{-10pt}\sum\limits_{(m,h)\in s_{P}}\hspace{-12pt}h$.
	\item \textbf{Recall}: The percentage of peaks in the measured spectrum that have a matching peak in the predicted spectrum: 
	$ 100 \times | s_{P} \cap s_{M} | \div |s_{M}| $.
\item \textbf{Precision}: The percentage of peaks in the predicted spectrum that have a matching peak in the measured spectrum: 
	$ 100 \times | s_{P} \cap s_{M} | \div |s_{P}| $.	
\item \textbf{Jaccard Score}:	
	$ | s_{P} \cap s_{M} | \div |s_{P} \cup s_{M}| $.
\end{enumerate} 

The intensity weighted metrics were included because the unweighted precision and recall values can be misleading in the presence of low-level noise --e.g. when there are many small peaks in the measured spectrum. The weighted metrics place a greater importance on matching higher intensity peaks, and therefore give a better indication of how much of a spectrum has been matched. However, these weighted metrics can also be susceptible to an over-emphasis of just one or two peaks, and in particular of the peak corresponding to the precursor ion. Consequently, we think it is informative to consider both weighted and non-weighted metrics for recall and precision.

\paragraph{Models for Comparison}:
The pre-existing methods, --e.g. MetFrag, FingerID -- do not output a predicted spectrum, but skip directly to metabolite identification. 
So, instead we compare against:
\begin{itemize}
\item \textbf{Full Enumeration}: This model considers the predicted spectrum to be one that enumerates all possible fragments in the molecule's fragmentation tree with uniform intensity values. 
\item \textbf{Heuristic} (tripeptides only): This model enumerates known peptide fragmentations as described by \citep{Papayannopoulos1995}, including $b_{n}$, $y_{n}$, $b_{n} - H_{2}O$, $y_{n}- H_{2}O$, $b_{n} - NH_{3}$, $y_{n}- NH_{3}$ and immonium ions.
\end{itemize}

\begin{figure*}
\begin{center}
\includegraphics[scale=0.48]{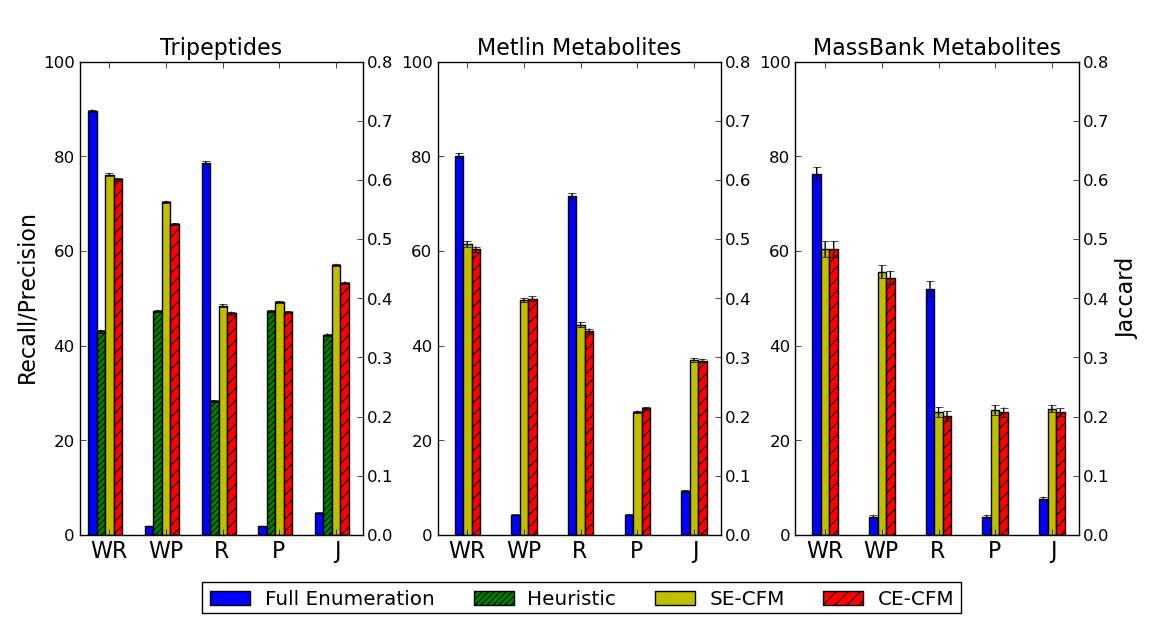}
\caption{Spectrum prediction results for tripeptides (left), metabolites from Metlin (middle) and metabolites from MassBank (right). The x-axes show the five metrics: Weighted Recall (WR), Weighted Precision (WP), Recall (R), Precision (P) and Jaccard (J), averaged across the three energy levels for each test molecule. Bars display mean scores $\pm$ standard error. In each plot, note that the y-axis for Jaccard (on right) is different from the others (on left).}
\label{fig:MetricsPeptides}
\end{center}
\end{figure*}

\paragraph{Results}:

The results are presented in Figure~\ref{fig:MetricsPeptides}.
%\begin{figure}
%\begin{center}
%\includegraphics[scale=0.44]{specpredict_results_metab.png}
%\caption{Spectrum prediction results for metabolites. Results for individual energy spectra are displayed on the left, and the averaged scores on the right. Bars display mean $\pm$ standard deviation.}
%\label{fig:MetricsMetab}
%\end{center}
%\end{figure}
For all three data sets, SE-CFM and CE-CFM obtain several orders of magnitude better precision and Jaccard scores than the full enumerations of possible peaks.
%These comparisons indicate the importance of considering the probability of each fragmentation, rather than just enumerating all explanations for a peak without regard for how likely those explanations are. 
%The results also show that our methods are able to model which peaks are more likely with some accuracy.
There is a corresponding loss of recall. However, if we take into account the intensity of the measured peaks, by considering the weighted recall scores, we see that our methods perform well on the more important, higher intensity peaks. 
More than 75\% of the total peak intensity in the tripeptide spectra, and approximately 60\% of the total peak intensity in the metabolite spectra, were predicted.

The results presented in Figure~\ref{fig:MetricsPeptides} show scores averaged across the three energy levels for each molecule. If we consider the results for the energy levels separately (see supplementary data), we find that the low and medium energy results are much better for all methods we assessed.
For example, in the case of the low energy spectra, the weighted recall scores for SE-CFM are 78\%, 73\% and 81\% for the tripeptide, Metlin metabolite and MassBank metabolite data sets respectively, as compared to 73\%, 29\% and 37\% respectively for the high energy spectra.
The poorer high energy spectra results may be due to increased noise and a lower predictability of events at the higher collision energies. 
Another possible explanation is that the even-electron rule and other assumptions listed in Section~\ref{StateSpace} may be less reliable when there is more energy in the system. \hl{Or perhaps it is simply a factor of the number of peaks per energy level, given that the median numbers of peaks in the measured and predicted spectra respectively were 5 and 6 in the low, 9 and 16 in the medium and 12 and 30 in the high energy spectra. }

In the case of the tripeptide data, our methods achieve higher recall scores and similar rates of precision to that of the heuristic model of known fragmentation mechanisms, resulting in improved Jaccard scores.
Since peptide fragmentation mechanisms are fairly well understood, this result is not intended to suggest that our method should be used in place of current peptide fragmentation programs, but rather to demonstrate that SE-CFM and CE-CFM are able to extract fragmentation patterns from data to a similar extent to human experts, given a sufficiently large and consistent data set. Like our methods, the heuristic models also perform better for the lower energy levels, with a weighted recall score of 66\% for the low energy, as compared to only 24\% for the high energy.

Unsurprisingly, being a smaller and more diverse data set, the Metlin metabolite results are poorer than those of the tripeptides. However the weighted recall for both our methods is still above 60\% and the precision and Jaccard scores are much higher than for the full enumeration, suggesting that the CFM model is still able to capture some of the common fragmentation trends. 

The weighted recall and precision results for the MassBank metabolites are fairly comparable to those of the Metlin metabolites. There is a small loss in the non-weighted recall, however this is probably due to a higher incidence of low-level noise in the MassBank data. This results in a small loss in the average Jaccard score. However these results demonstrate that the fragmentation trends learned still apply to a significant degree on data collected at a different time in a different laboratory.

%For all three datasets, we can observe a trend in the results for the different energy levels. The low and medium energy results are better for all methods %trialled. The poorer high spectrum results may be due to more noise and less predictable events at the higher energy level. Another possible explanation is that the even-electron rule and other assumptions listed in Section~\ref{StateSpace} may be less reliable when there is more energy in the system. %However, given that CE-CFM and SE-CFM still out-perform the full enumeration in precision and Jaccard, some predictability in fragmentations appears to be captured in the models.

%The above results only indirectly consider the intensity values of the predicted peaks, via the thresholding of low probability peaks. \
Since this is the first method, to the authors' knowledge, capable of predicting intensity values as well as m/z values, we also investigated the accuracy of CFM's predicted intensity values. We found that the Pearson correlation coefficients for matched pairs of predicted and measured peaks, were 0.7, 0.6 and 0.45 for the low, medium and high spectra respectively (SE-CFM and CE-CFM results were not significantly different). This indicates a positive, though imperfect correlation. Full results and scatter plots are contained in the supplementary data.

\hl{Running on a 2.2GHz Intel Core i7 processor, the median run-time for the spectrum predictions for each molecule in the Metlin metabolite data set was 5 seconds. Larger molecules with more ring systems generally take longer as they have so many more fragmentation possibilities in the initial enumeration. For molecules with no rings, the median run-time was 2 seconds, whereas for molecules with 3 or more rings, the median run-time was 9 seconds. The longest run-time in the Metlin metabolite set was for Troleandomycin (Metlin ID 41012), which has a molecular weight over 800~Da and contains three ring systems, one of which is size 14. It took just under 5 minutes.}

\subsection{Metabolite Identification}
\label{sec:MetabId}

Here we apply our CFM MS/MS spectrum predictions to a metabolite identification task. 
For each molecule, we produce two candidate sets via queries to two public databases of chemical entities:
\begin{enumerate}
\item We query the PubChem compound database \citep{Bolton2008} for all molecules within 5~ppm of the known molecule mass. This simulates the case where little is known about the candidate compound, but the parent ion mass is known with high accuracy.
\item We query KEGG (Kyoto Encyclopedia of Genes and Genomes) \citep{Kanehisa2006} for all the molecules within 0.5~Da of the known molecular mass. This simulates the case where the molecule is thought to be a naturally occurring metabolite, but there is more uncertainty in the target mass range.
\end{enumerate} 

To conduct this assessment, duplicate candidates were filtered out --i.e. those with the same chemical structure, including those that only differ in their stereochemistry.
Charged molecules and ionic compounds were also removed since the program assumes single fragment, neutral candidates (to which it will add a proton).
After filtering, the median number of candidates returned from PubChem was 911 for the tripeptides and 1025 for the metabolites. 
Note that 9 tripeptides and 57 of the Metlin metabolites were excluded from this testing because no matching entry was found in PubChem for these molecules.
The KEGG queries were only carried out for the metabolite data. 
The median number of candidates returned was 22, however no matching entry was found in KEGG for 833 of the Metlin metabolites and 111 of the MassBank metabolites.

Whenever a matching entry could be found, we ranked the candidates according to how well their predicted low, medium and high spectra matched the measured spectra of the test molecule.
The ranking score we used was the Jaccard score described in Section~\ref{sec:SpectrumPrediction}.

\begin{figure*}
\begin{center}
\includegraphics[scale=0.48]{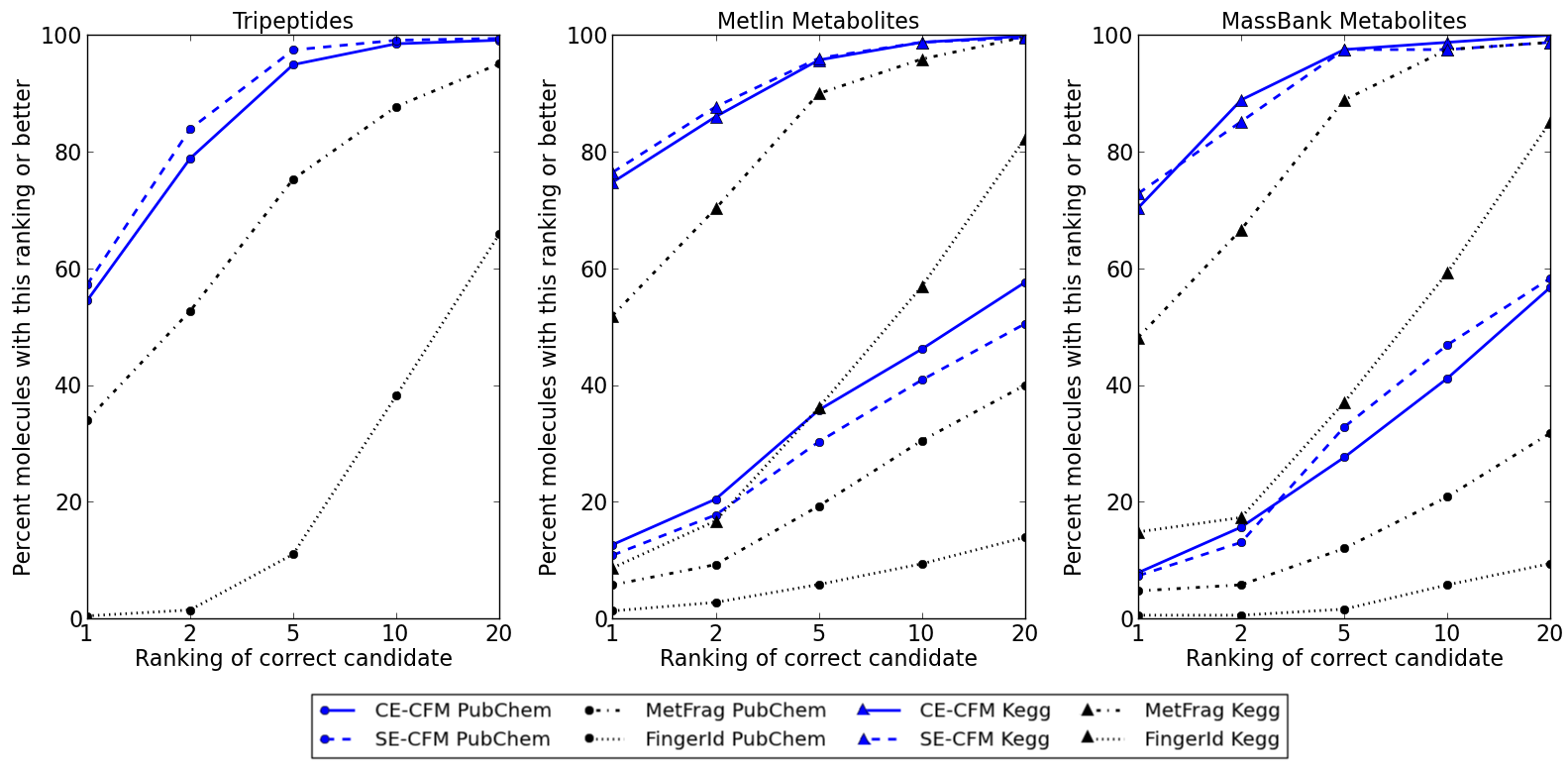}
\caption{Ranking results for metabolite identification, comparing both CFM variants with MetFrag and FingerID for tripeptides (left), metabolites from Metlin (middle) and validation metabolites from MassBank (right), querying against PubChem within 5~ppm (circles) and KEGG within 0.5~Da (triangles). Note that our methods out-perform both MetFrag and FingerID on all metrics, regardless of the database used.}
\label{fig:rankingresults}
\end{center}
\vspace{-1.25em}
\end{figure*}

We compared the ranking performance of our SE-CFM and CE-CFM methods against those of MetFrag~\citep{Wolf2010} and FingerID~\citep{Heinonen2012}. 
We used the same candidate lists for all programs. 
For candidate molecules with equal scores, we had each program break ties in a uniformly random manner. This was in contrast to the original MetFrag code, which used the most pessimistic ranking; we did not use that approach as it seemed unnecessarily pessimistic. 
We set the mass tolerances used by MetFrag when matching peaks to the same as those used in our method (maximum of 0.01Da and 10ppm). 
MetFrag and FingerID only accept one spectrum, so to input the three spectra we first merged them as described by~\citep{Wolf2010}: 
we took the union of all peaks, and then merge together any peaks within 10~ppm or 0.01~Da of one another, retaining the average mass and the maximum intensity of the two. 
In FingerID we used the linear High Resolution Mass Kernel including both peaks and neutral losses, and trained using the same cross-fold sets as for our own method.
Overall, we attempted to assess CFM, MetFrag and FingerID as fairly as possible, using identical constraints, identical databases and near-identical data input. 
The results are shown in Figure~\ref{fig:rankingresults}. 

As seen in this figure, our CFM method achieved substantially better rankings than both the existing methods on all three data sets, for both the PubChem and KEGG queries. When querying against KEGG, our methods found the correct metabolite as the top-scoring candidate in over 70\% of cases for both metabolite sets and almost always $(>95\%)$ ranked the correct candidate in the top 5. In comparison, MetFrag ranked the correct metabolite first in approximately 50\% of cases for both metabolite sets, and in the top 5 in 89\%. FingerID ranked the correct metabolite first in less than 15\% of cases.

For PubChem, our methods performed well on the tripeptide data, identifying the correct metabolite as the top-scoring candidate in more than 50\% of cases and ranking the correct candidate in the top 10 for more than 98\% of cases. This is again convincingly better than both MetFrag and FingerId, which rank the correct candidate first in less than 35\% and 2\% of cases respectively.

For the metabolite data, CE-CFM and SE-CFM were able to identify the correct metabolite in only 12\% and 10\% of cases respectively, however given that this is from a list of approximately one thousand candidates, this performance is still not bad. Once again, it is substantially better than MetFrag and FingerID, which correctly identified less than 6\% and 1\% of cases respectively. Our methods rank the correct candidate in the top 10 in more than 40\% of cases on both data sets, as compared to MetFrag's performance of 31\% on the Metlin metabolites and 21\% on the MassBank metabolites. Additionally, the top-ranked compound was found to have the correct molecular formula in more than 88\% of cases for SE-CFM and 90\% of cases for CE-CFM, suggesting that both methods mainly fail to distinguish between isomers.  While the performance of all three methods (CFM, MetFrag and FingerID) is not particularly impressive for the PubChem data sets (i.e. $<$12\% correct) we would argue that the PubChem database is generally a poor database choice for anyone wishing to do MS/MS metabolomic studies.  With only 1\% of its molecules having a biological or natural product origin, one is already dealing with a rather significant challenge of how to eliminate a 100:1 excess of false positives.  So we would regard the results from the PubChem assessment as a "worst-case" scenario and the results from the KEGG assessment as a more typical metabolomics scenario.

The results for CE-CFM showed minimal difference when compared to those of SE-CFM, casting doubt on whether the additional complexity of CE-CFM is justified. 
However we think this idea is still interesting as a means for integrating information across energy levels and may yet prove more useful in future work.  

\hl{The running time of the metabolite identifications is mainly dependent on the number of candidate molecules and the time taken to predict the spectra for each. For example, taking 1000 candidates (as in the PubChem tests) at the median spectrum prediction run-time of 5~seconds} (see Section~\ref{sec:SpectrumPrediction}), \hl{the identification would be expected to take in the order of 1.5~hours. Taking only 22 candidates (as in the KEGG tests), this reduces to 2~minutes. It would be trivial to parallelize the computation by distributing candidates across processors. When repeatedly querying against the same database, it may also be expedient to precompute the predicted spectra to reduce the identification run-time. For example, our web server interface} \url{http://cfmid.wishartlab.com} \hl{provides access to precomputed spectra for all 40,000 compounds in HMDB and over 10,000 compounds in KEGG.} We encourage readers to make use of \hl{this web server, as well as} our executables and source code, made available at \url{http://sourceforge.net/projects/cfm-id/}.

%It is possible that there is insufficient information in the MS/MS data to differentiate between some candidates.
%However, the higher performance obtained on the tripeptide data gives us hope that the use of more extensive training datasets may improve the results for a wider range of metabolites. 
%Further gains may also be made via better chemical feature representations, or by further narrowing the candidate space using prior knowledge. 

\section{Conclusion}

We have proposed a model for the ESI-MS/MS fragmentation process and a method for training this model from data.
The performance has been benchmarked in cross validation testing on a large molecule set, and further validated using an additional dataset from another laboratory. 
Head-to-head comparisons using multiple data sets under multiple conditions show that the CFM method significantly outperforms existing state-of-the-art methods, and has attained a level that could be useful to experimentalists performing metabolomics studies.

\end{document}